\documentclass{ifacconf}
\usepackage[fleqn]{amsmath}

\DeclareMathOperator{\diag}{diag}                                                           

\usepackage{mathrsfs,natbib}										
\usepackage{amsmath, amssymb,amsfonts,dsfont}
\usepackage{etoolbox}

\usepackage{theorem}
\newtheorem{theorem}{Theorem}                						
\newtheorem{lemma}{Lemma}
\newtheorem{definition}{Definition}
               						
\newtheorem{remark}{Remark}

\newtheorem{example}{Example}

\usepackage{mathtools}

\usepackage{graphics} 										
\usepackage{epsfig} 													

								\renewcommand{\vec}{\boldsymbol}								
\renewcommand{\thefootnote}{\fnsymbol{footnote}}					\DeclareMathAlphabet{\mathpzc}{OT1}{pzc}{m}{it}
\DeclareMathAlphabet{\mathcal}{OMS}{cmsy}{m}{n}

\usepackage{balance}
\usepackage[american,cute inductors,smartlabels]{circuitikz}
\usepackage{pgf}
\usepackage{tikz}
\usepackage{tabularx}
\usepackage{longtable}
\usepackage{booktabs}
\usetikzlibrary{arrows,automata}

\usepackage{textcomp}

\newcommand{\blue}[1]{\textcolor{black}{#1}}

\usepackage{thmtools}
\newtheorem{corollary}[theorem]{Corollary}
\newtheorem{proposition}[theorem]{Proposition}

\newcommand{\cE} { {\cal E}}
\newcommand{\bR} { {\mathbb R}}
\begin{document}
	\begin{frontmatter}
		
\title{\LARGE \bf Krasovskii's Passivity} 
		
		%\thanks[footnoteinfo]{Sponsor and financial support acknowledgment
		%	goes here. Paper titles should be written in uppercase and lowercase
		%	letters, not all uppercase.}
		
		\author[First]{Krishna C. Kosaraju} 
		\author[JP]{Yu Kawano} 
		\author[First]{Jacquelien M.A. Scherpen}
		
		\address[First]{Jan C. Wilems Center for Systems and Control, ENTEG, Faculty of Science and Engineering, University of Groningen, Nijenborgh 4, 9747 AG Groningen, the Netherlands 
			{\tt\small \{k.c.kosaraju, j.m.a.scherpen\}@rug.nl}.}
			\address[JP]{Faculty of Engineering, Hiroshima University, Kagamiyama 1-4-1, Higashi-Hiroshima 739-8527, Japan {\tt\small ykawano@hiroshima-u.ac.jp}.}
			\thanks{	This work is supported by the Netherlands Organisation for Scientific Research through
		Research Programme ENBARK+ under Project 408.urs+.16.005.}
		
\begin{abstract}                % Abstract of not more than 250 words.
In this paper we introduce a new notion of passivity which we call Krasovskii's passivity and provide a sufficient condition for a system to be Krasovskii's passive. Based on this condition, we investigate classes of port-Hamiltonian and gradient systems which are Krasovskii's passive. Moreover, we provide a new interconnection based control technique based on Krasovskii's passivity. Our proposed control technique can be used even in the case when it is not clear how to construct the standard passivity based controller, which is demonstrated by examples of a Boost converter and a parallel RLC circuit. 
\end{abstract}
		
		
	\end{frontmatter}
\maketitle
\thispagestyle{empty}
\pagestyle{empty}

\section{INTRODUCTION}\vspace{0mm}
The applications of passivity, or more generally dissipativity, are ubiquitous in various examples in systems theory, such as, stability, control, robustness; see, e.g.,~\cite{915398,willems1972dissipative,van2000l2,Khalil:96}. However, since passivity depends on the considered input-output maps and the classical notion does not always suffice, variations of passivity concepts are still being developed, such as the differential passivity in~\cite{6760930,VANDERSCHAFT201321} and the counter clockwise concept in~\cite{angeli2006systems}. 

One of the standard tool for discovering passive input-output maps is by construction of energy-like functions, the so called storage functions. Several methods/frameworks have been proposed and developed in a quest to find new storage functions, such as the port-Hamiltonian (pH) systems theory, the Brayton-Moser (BM) framework, and different variants of Hill-Moylan's lemma, \cite{hill1980dissipative}. However, there is no universal way for constructing storage functions, which coincides with the difficulty of finding a Lyapunov function for stability analysis.

%For the construction of a Lyapunov function, there is a technique called Krasovskii's method found in~\cite{Khalil:96}. The main idea of this approach is to employ a quadratic function of the vector field as a Lyapunov function. \cite{ECC} originally applies this construction for the storage function of electrical circuits. Similar ideas are found for specific control problems for electrical circuit and the convex optimization in~\cite{kosaraju2017control,2018arXiv181102838C,kosaraju2018stability,kosaraju2018differential}. 
\blue{For the construction of a Lyapunov function, there is a technique called Krasovskii's method found in~\cite{Khalil:96}. The main idea of this approach is to employ a quadratic function of the vector field as a Lyapunov function. In  ~\cite{kosaraju2017control,2018arXiv181102838C,kosaraju2018stability,kosaraju2018differential} the authors explored the idea of using this as a storage function for presenting new passivity properties of systems such as electrical networks, primal-dual dynamics, and HVAC systems. More recently, in \cite{ECC} the authors partially formalized this idea - in application towards constant power-loads.}
An interesting fact is that this type of storage functions is helpful for stabilization problems of an electrical circuit for which finding a suitable storage function is difficult. Although its utility is demonstrated by aforementioned references, the passivity property with Krasovskii's types storage function, which we name Krasovskii's passivity has not been defined for general nonlinear systems. As a consequence, properties and structures of general Krasovskii's passive systems have not been investigated. The aim of this paper is to develop Krasovskii's passivity theory and to provide a first perspective on its use for control.

We list the main contribution below:
\begin{itemize}
	\item[(i)] We formally define Krasovskii's passivity and provide a sufficient condition. Moreover, we show that Krasovskii's passivity is preserved under feedback interconnection. 
	\item[(ii)] We present sufficient conditions for port-Hamiltonian and Brayton-Moser systems to be Krasovskii's passive. Moreover, in certain cases we also construct Krasovskii's storage function. We also present a brief introduction on designing controllers using Krasovskii's passivity. 
\end{itemize}
{\em Outline}: This paper is outlined as follows. Section II gives a motivating example. Section III presents the definition of Krasovskii's passivity and its properties. Section IV provides applications of Krasovskii's passivity including a novel control technique. Throughout the paper, we illustrate our findings using parallel RLC and Boost converter systems as examples.

{\it Notation:}
The set of real numbers and non-negative real numbers are denoted by $\bR$ and $\bR_+$, respectively. For a vector $x \in \bR^n$ and a symmetric and positive semidefinite matrix~$M \in \bR^{n \times n}$, define~$\| x\|_M:= (x^\top M x)^{1/2}$. If~$M$ is the identity matrix, this is nothing but the Euclidean norm and is simply denoted by~$\|x\|$. For symmetric matrices $P,Q \in \bR^{n\times n}$, $P \le Q$ implies that $Q-P$ is positive semidefinite.
\vspace{-3mm}
\section{Motivating Examples}\vspace{0mm}
In this section, we first recall the definition of passivity for the nonlinear system. Then, we present an example to explain the motivation for introducing a novel passivity concept from the controller design point of view. Consider the following continuous time input-affine nonlinear system:\vspace{0mm}
\begin{align}\label{eq:sys}
\Sigma: \dot{x}=f(x,u):=g_0(x)+\sum_{i=1}^{m}g_i(x)u_i,
\end{align}
where~$x:\bR \to \bR^n$ and~$u=[\begin{array}{ccc}u_1 & \dots & u_m\end{array}]^\top:\bR \to  \bR^m$ denote the state and input, respectively. Functions~$g_i:\bR^n\to\bR^n$, $i=0,1,\dots ,m$ are of class~$C^1$, and define~$g:=[\begin{array}{ccc}g_1 &\dots & g_m\end{array}]$ by using the latter $m$ vector fields. Throughout this paper, we assume that the system~\eqref{eq:sys} has at least one forced equilibrium. Namely, we assume that the following set
\begin{align}\label{set:equilibrium}
\cE :=\{(x^*,u^*) \in \bR^n \times \bR^m:f(x^*,u^*)=0\}
\end{align}
is not empty.

Passivity is a specific dissipativity property. % and our novel passivity concept is also defined as a specific dissipativity with a specific storage function. 
For self-containedness, we show the definitions of dissipativity, passivity and its variations for system~\eqref{eq:sys}.
\begin{definition}(\cite{willems1972dissipative,van2000l2,Khalil:96})\label{def:dp}
The system~\eqref{eq:sys} is said to be dissipative with respect to a supply rate~$w:\bR^n \times \bR^m \to \bR$ if there exists a class~$C^1$ storage function~$S:\bR^n\to \bR_{+}$ such that
\begin{align}\label{eq_dissipative}
\frac{\partial S(x)}{\partial x} f(x,u)  \le w(x,u)
\end{align}
for all~$(x,u) \in \bR^n \times \bR^m$.
\end{definition}
\begin{definition}[Equilibrium Independent Dissipativity]\label{def:EID}
The system~\eqref{eq:sys} is said to be equilibrium independent dissipative (EID) with respect to a supply rate~$w:\bR^n \times \bR^m \to \bR$ if this is dissipative in the sense of Definition~\ref{def:dp}, and for every forced equilibrium $(x^*,u^\ast)\in \cE$, $S(x^*)=0$ and $w(x^*,u^*)=0$ hold.
\end{definition}

Compared with EID in \cite{simpson2018equilibrium}, the above definition is more general in the sense that we do not assume that the supply rate is of the structure $w(u-u^*,y-y^*)$. If system~\eqref{eq:sys} is dissipative or EID with respect to a supply rate $w= u^\top y$, then it is said to be passive; see e.g.~\cite{van2000l2,Khalil:96}. Especially, we call this passivity the standard passivity to distinguish with the new passivity concept provided in this paper. 

Passivity is characterized by a suitable storage function, and it is typically the total energy of the system. However, as demonstrated by the following example, passivity with the total energy as a storage function is not always helpful for analysis and controller design.
\begin{example}\label{eg:boost}
We consider the average governing dynamic equations of the Boost converter; see e.g.~\cite{ECC,2018arXiv181102838C, jeltsema2004tuning} for more details about this type of models. Its state space equation is given by
\begin{align}
\begin{split}
\label{eq:boost}
-L\dot{I}(t) &=R I(t)+(1 - u(t)) V(t) - V_s,\\
C\dot{V}(t) &=(1 - u(t)) I(t) - G V(t),
\end{split}
\end{align}
where $L,R,V_s,C,G$ are positive constants, $I,V: \bR \to \bR$ are the state variables (average current and voltage), and  $u:\bR \to \bR$ is the control input (duty ratio, $u\in [0,1]$). Its total energy is\vspace{0mm}
\begin{align*}
S(I,V) = \frac{1}{2} (L I^2 + C V^2).
\end{align*}
Its time derivative along the trajectory of the system is computed as
\begin{align*}
\frac{d}{dt}S(I,V) = - R I^2 - G V^2 + V_s I \leq  V_s I.
\end{align*}
This implies the boost converter is passive with respective to the source voltage $V_s$ and the current $I$. However, $V_s$ is a constant and cannot be controlled. Therefore, the standard passivity with the total energy~$S(I,V)$ is not helpful for designing the control input~$u$.% To overcome this, in \cite{ortega2013passivity}, the authors considered incremental energy (Bregman divergence of $S(I,V)$) as the the storage function.

In order to address this issue, recently, \cite{2018arXiv181102838C, ECC} provides a new passivity based control technique by using the following extended system~[\cite{van1982observability}]:
\begin{align}
\begin{split}
\label{eq:boost2}
-L\dot{I}(t) &=R I(t)+(1 - u(t)) V(t) - V_s,\\
C\dot{V}(t) &=(1 - u(t)) I(t) - G V(t),\\
\dot u(t) &= u_d (t),
\end{split}
\end{align}
where $I,V,u:\bR\to\bR$ are new state variables, and $u_d:\bR \to \bR$ is a new input variable. Note that a forced equilibrium point $(x^*,u^*)$ of the boost converter~\eqref{eq:boost} is an equilibrium of its extended system~\eqref{eq:boost2}. The extended system is dissipative with respect to $u_d^\top (\dot{I}V-\dot{V}I)$ with the following storage function,
\begin{align}
\bar S(\dot I, \dot V) = \frac{1}{2} (L \dot I^2 + C \dot V^2). \label{eq:storage_boost2}
\end{align}
Moreover, the following feedback control input
\begin{align}
u_d  = K(u -u^*) - (\dot{I}V-\dot{V}I), \ K>0, \label{eq:con_ex}
\end{align}
stabilizes the equilibrium $(x^*,u^*)$, see \cite{2018arXiv181102838C, ECC}. \blue{Note that the dynamic controller \eqref{eq:con_ex} is different from the one presented in \cite{ortega2013passivity}, where the authors use damping injection technique which results in a local stabilizing controller.}
\end{example}

Although the extended system~\eqref{eq:boost2} and its storage function~\eqref{eq:storage_boost2} help the controller design, the interpretation of them has not been provided yet. In this paper, our aim is to understand the above new control technique by developing passivity theory for the input-affine nonlinear system~\eqref{eq:sys}.\vspace{0mm}
\section{Krasovskii's passivity}\vspace{0mm}
\subsection{Definition and Basic Properties}\vspace{0mm}
In this subsection, we investigate a novel passivity concept motivated by Example~\ref{eg:boost}. First, we provide the definition and then show a sufficient condition. 

In Example~\ref{eg:boost}, the new storage function~$\bar S(\dot I, \dot V)$ for the extended system~\eqref{eq:boost2} is employed by considering input port variables $u_d$ instead of $u$. The structure of this storage function is similar to the Lyapunov function constructed by Krasovskii's method~\citep{Khalil:96} for the autonomous system~$\dot x = g_0(x)$, namely~$V(x)=\|g_0^\top (x)\|_Q/2$ for positive definite~$Q$. We focus on this type of specific storage functions and use it for analysis of the following extended system of~\eqref{eq:sys}:
\begin{align}\label{eq:sys_ext}
\left\{\begin{array}{l}
\dot x=f(x,u),\\
\dot u = u_d,
\end{array}\right.
\end{align}
where~$[\begin{array}{cc}x^\top & u^\top\end{array}]^\top :\bR \to \bR^{n+m}$ and~$u_d:\bR \to \bR^m$ are the new states and inputs. 

Now, we are ready to define a novel passivity concept. From its structure, we call it Krasovskii's passivity.
\begin{definition}[Krasovskii's passivity]\label{def:KP}
Let~$h_K:\bR^n \times \bR^m\to \bR^m$. Then, the nonlinear system~\eqref{eq:sys} is said to be Krasovskii passive if its extended system~\eqref{eq:sys_ext} is dissipative with respect to the supply rate~$u_d^\top h_K(x,u)$ with a storage function~$S_K(x,u)=(1/2)\|f (x,u)\|_Q^2$, where $Q \in \bR^{n\times n}$ is symmetric and positive semidefinite for each~$(x,u) \in \bR^n \times \bR^m$.
\end{definition}

Note that different from the Lyapunov function, the storage function can be positive semidefinite. However, when one designs a stabilizing controller based on Krasovskii's passivity, the storage function~$S_K(x,u)$ is used as a Lyapunov candidate, and thus $Q$ is chosen as positive definite.
\begin{remark}
In Example~\ref{eg:boost}, for the sake of simplicity of notation, we describe the storage function as a function of $(\dot I, \dot V)$. However, when we proceed with analysis, we use the function $S_K(I,V,u)$ in the form of Definition~\ref{def:KP} instead. 
%The main reason why we use the function of $(I, V)$ is to conclude the property of the system whose state variables are $(I, V)$ but not $(\dot I, \dot V)$.
\end{remark}

\begin{remark}
In contraction analysis with constant metric, a Lyapunov function constructed by Krasovskii's method is found in~\cite{6632882}. Contraction analysis is based on the variational system along the trajectory of the system~\eqref{eq:sys},
\begin{align}\label{eq:vsys}
\frac{d\delta x}{dt}=\frac{\partial f(x,u)}{\partial x} \delta x + \frac{\partial f(x,u)}{\partial u} \delta u. 
\end{align}
As a passivity property of the variational system, differential passivity is proposed by~\cite{6760930,VANDERSCHAFT201321}. In the constant metric case, the corresponding storage function is in the form $\| \delta x\|_Q^2$. One notices that $\delta x=f(x,u)$ and $\delta u =u_d$ satisfies the dynamics of~\eqref{eq:vsys}. Therefore, one can employ the results on differential passivity for the analysis of Krasovskii's passivity. It is worth emphasizing that differential passivity is generally not used for controller design, since it is not always easy to connect the control input of the variational system~$\delta u$ with the original system.
\end{remark}

In the following proposition, we confirm that the following sufficient condition for differential passivity in~\cite{VANDERSCHAFT201321} is also a sufficient condition for Krasovskii's passivity.
\begin{proposition}\label{prop:Krasovskii_passivity}
Let~$Q\in \bR^{n\times n}$ be a symmetric and positive semidefinite matrix. If the vector fields $g_0$, $g_i$, $i\in \{1\cdots m\}$ of the system~\eqref{eq:sys} satisfies
	\begin{align}
	&Q_{g_0}(x):=Q\dfrac{\partial g_0(x)}{\partial x}+\dfrac{\partial^\top g_0(x)}{\partial x} Q\leq 0,\label{eq1:differential_passivity}
	\\
	&Q_{g_i}(x):=Q\dfrac{\partial g_i(x)}{\partial x}+\dfrac{\partial^\top g_i(x)}{\partial x}Q= 0,~\forall i=1,\dots,m, \label{eq2:differential_passivity}
	\end{align}
	then the system~\eqref{eq:sys} is Krasovskii passive for~$h_K(x,u):=g^\top (x)Q f(x,u)$.
\end{proposition}
\begin{pf}
Compute the Lie derivative of the storage function~$(1/2) \|f (x,u)\|_Q^2$ along the vector field of~\eqref{eq:sys_ext} as follows 
	\begin{align*}
&\dfrac{1}{2}\dfrac{d}{dt}\|f (x,u)\|_Q^2\\
&=f^\top(x,u) \left(Q_{g_0}(x)+\sum_{i=1}^{m}Q_{g_i}(x)u_i\right)f(x,u)\\
&+u_d^\top g^\top (x)Q f(x,u) \leq  u_d^\top h_K (x,u).
\end{align*}
That completes the proof.\qed
\end{pf}

It easily follows that a system satisfying the conditions in Proposition~\ref{prop:Krasovskii_passivity} is EID. If one evaluates the value of the storage function and supply-rate on the set $\cE$ (set of all feasible forced equilibria), we get 
\begin{align}\label{eq_EID_kras}
&(1/2)\|f (x,u)\|_Q^2=0, \\
&u_d^\top g^\top M f(x,u)=0,~\forall (x,u)\in \cE.
\end{align}
Therefore, the results on EID \cite{simpson2018equilibrium} can be applied for the analysis of Krasovskii's passive system.

As an application of Proposition~\ref{prop:Krasovskii_passivity}, we study the interconnection of Krasovskii passive systems. It is well-known that feedback interconnection of two standard passive systems is again a standard passive system. This property plays a crucial role in modeling, development of  controllers and robustness analysis. For the feedback interconnection of Krasovskii's passive system, we have a similar result.
\begin{proposition}\label{interconnection_kras}
	Consider two Krasovskii's passive systems $\Sigma_i$ (of the form \eqref{eq:sys} with states $x_i$ and inputs $u_i\in \bR^m$)  with respect to supply-rates $u_{d_i}^\top h_{K_i}(x_i,u_i)$ and Krasovskii's storage functions $S_{K_i}(x_i,u_i)$. Then the interconnection of two Krasovskii's  passive systems $\Sigma_1$ and $\Sigma_2$, via the following interconnection constraints
		\begin{align}\label{interconnection_rule}
		\begin{bmatrix}
		u_{d_1}\\u_{d_2}
		\end{bmatrix}=\begin{bmatrix}
		0&-1\\1&0
		\end{bmatrix}\begin{bmatrix}
		h_{K_1}\\h_{K_2}
		\end{bmatrix}+\begin{bmatrix}
		e_{d_1}\\e_{d_2}
		\end{bmatrix}
		\end{align}
	        is Krasovskii's passive with respect  to a supply-rate $e_{d_1}^\top h_{K_1}+e_{d_2}^\top h_{K_2}$ and Krasovskii's storage function $S_{K_1}+S_{K_2}$, where $\dot e_1=e_{d_1}$, $\dot e_2=e_{d_2}$ and $e_1, e_2:\mathbb{R}\rightarrow\mathbb{R}^m$ are external inputs.
\end{proposition}
\begin{pf}
	For $i\in \{1,2\}$ the systems $\Sigma_i$ satisfy
	%\begin{eqnarray}
$	\dot S_{K_i} \leq u_{d_i}^\top h_{K_i}(x_i)$.
	%\end{eqnarray}
	Now consider the Lie derivative of $S_K=S_{K_1}+S_{K_1}$ along the vector fields of $\Sigma_1$, $\Sigma_2$, $\dot{u}_1=u_{d_1}$,~and $\dot{u}_2=u_{d_2}$
	\begin{align*}
		\dot{S}_K\leq u_{d_1}^\top h_{K_1}+u_{d_2}^\top h_{K_2}
		=e_{d_1}^\top h_{K_1}+e_{d_2}^\top h_{K_2}.\qed
	\end{align*}
%	In the second line, we use \eqref{interconnection_rule}.\hfill \qed%the interconnection constraints \eqref{interconnection_rule}.\hfill \qed
\end{pf}
\vspace{0mm}
\subsection{Port-Hamiltonian Systems}\vspace{0mm}
In this and next subsections, we investigate a general representation of Krasovskii passive systems. First, we consider port-Hamiltonian systems (PHSs). Various passive physical systems can be modelled as PHSs. Since Krasovskii passivity is a kind of passivity property, it is reasonable to study when PHSs become Krasovskii passive. 

From the structure of Krasovskii's passivity, we consider a different class from the standard PHS found in~\cite[Chapter~2.13]{sira2006control} whose interconnection matrix is also a function of input,\vspace{0mm}
\begin{eqnarray}\label{eq:ph_sys2}
	\dot{x}=\bar f(x,u):=\left( J_0+\sum_{i=1}^{m}J_i u_i-R\right)\dfrac{\partial H}{\partial x}(x) + G u_s
\end{eqnarray}
where the scalar valued function $H:\bR^n \to \bR_+$ is called a Hamiltonian, $u_s\in \bR^q$ is a constant vector, and $J_i \in \bR^{n \times n}$,~$\forall i\in \{0\cdots m\}$, $R\in \bR^{n \times n}$ and $G\in \mathbb{R}^{n\times q}$ are constant matrices. Moreover, $J_i \in \bR^{n \times n}$,~$\forall i\in \{0\cdots m\}$ and $R\in \bR^{n \times n}$ are
skew-symmetric and positive semidefinite, respectively. A PHS is Krasovskii passive under the following conditions obtained based on Proposition~\ref{prop:Krasovskii_passivity}.
\begin{corollary}\label{prop:Krasovskii_passivity_ham2}
	Consider a PHS \eqref{eq:ph_sys2}. Let~$Q\in \bR^{n\times n}$ be a symmetric and positive semidefinite matrix. The following statements hold.
	\begin{itemize}
		\item[(i)] If there exists a positive semidefinite $Q$ satisfying 
		\begin{align}\label{eq:ham_kras_pass_cond2}
		&Q(J_0-R)\dfrac{\partial^2 H}{\partial x^2}+\dfrac{\partial^2 H}{\partial x^2}(-J_0-R)Q\leq 0,\\
		&QJ_i\dfrac{\partial^2 H}{\partial x^2}-\dfrac{\partial^2 H}{\partial x^2}J_iQ=0,~\forall i\in \{1\cdots m\}\label{eq:ham_kras_pass_cond3}
		\end{align}
		then the PHS is Krasovskii passive for~$h_K:=\bar g^\top (x) Q \bar f(x,u)$, where the columns of $\bar g$ are given by $J_i\frac{\partial H}{\partial x}$. 
		\item[(ii)] Furthermore if $\frac{\partial^2 H}{\partial x^2}$ is constant, then $Q:=\alpha\frac{\partial^2 H}{\partial x^2}$ satisfies equations  \eqref{eq:ham_kras_pass_cond2} and \eqref{eq:ham_kras_pass_cond3} for all $\alpha>0$.
	\end{itemize}
\end{corollary}
\begin{pf}
The statement (i) follows from Proposition~\ref{prop:Krasovskii_passivity} by choosing $g_0=(J_0-R)\dfrac{\partial H}{\partial x}+Gu_s$ and $g_i=J_i\dfrac{\partial H}{\partial x}$. The statement (ii) can readily be confirmed. \qed
\end{pf}

In fact, the boost converter can be represented as a PHS satisfying the conditions in Corollary~\ref{prop:Krasovskii_passivity_ham2}.
\begin{example}[Revisit of Example \ref{eg:boost}]
Consider the boost converter in Example \ref{eg:boost}, which takes the pH form given in \eqref{eq:ph_sys2} as follows
\begin{align}\label{eq:boost_pH}
	&\hspace{-5mm}\begin{bmatrix}
	\dot{I}\\ \dot{V}
	\end{bmatrix}=\left(\begin{bmatrix}
	0&-\frac{u}{LC}\\ \frac{u}{LC}&0
	\end{bmatrix}-\begin{bmatrix}
	\frac{R}{L^2}&0\\0&\frac{G}{C^2}
	\end{bmatrix}\right) \begin{bmatrix}
	\frac{\partial H}{\partial I}\\\frac{\partial H}{\partial V}
	\end{bmatrix}-\begin{bmatrix}
	\frac{1}{L}\\0
	\end{bmatrix}V_s,\\
	&\hspace{-5mm}H=\frac{1}{2}LI^2+\frac{1}{2}CV^2.
\end{align}
Note that $\frac{\partial^2 H}{\partial x^2}=\diag \{L,C\}$ is constant. According to~Proposition \ref{prop:Krasovskii_passivity_ham2}-(ii), the boost converter is Krasoskii passive with the storage function $S_K(x,u)=(1/2)\|f (x,u)\|_Q^2$, where $Q=\frac{\partial^2 H}{\partial x^2}$. This coincide with Example \ref{eg:boost}.
\end{example}

\subsection{Gradient Systems}
Similar to PHSs, gradient systems, see e.g.~\cite{cortes2005characterization} arise from physics as well. In general, there is no direct connection between these two types of systems except when the gradient system is passive~[\cite{Schaft:11}]. In this subsection, we investigate when gradient systems become Krasovskii passive.

A gradient system~\citep{cortes2005characterization} is given as follows,
\begin{eqnarray}\label{BM_gradient_form}
	D\dot{x}=\tilde f(x,u):=\dfrac{\partial P(x)}{\partial x}+B(x) u,
\end{eqnarray}
where $P:\bR^n \to \bR$ is a scalar valued function called a potential function, $D \in \bR^{n \times n}$ is a nonsingular and symmetric matrix called a pseudo metric, and $B \in  \bR^n \times \bR^m$.

A gradient system is Krasovskii passive under the following conditions obtained based on Proposition~\ref{prop:Krasovskii_passivity}. Since the proof is similar as that for Proposition~\ref{prop:Krasovskii_passivity_ham2}, it is omitted.
\begin{proposition}\label{prop:Krasovskii_passivity_BM}
Consider a gradient system \eqref{BM_gradient_form}. If there exist a symmetric and positive semidefinite matrix $M$ satisfying 
		\begin{align}\label{eq:BM_kras_pass_cond}
		&D M  \dfrac{\partial^2 P }{\partial x^2}+\dfrac{\partial^2 P }{\partial x^2} M D\leq 0
		\end{align}
		then the gradient system is Krasovskii passive with respect to~$h_K:=B^\top M\tilde f(x,u)$, where~$Q :=DMD$. 
\end{proposition}

We provide an electrical circuit that can be represented as a gradient systems satisfying the condition in Proposition~\ref{prop:Krasovskii_passivity_BM}.
\begin{example}\label{eg:RLCdyn_ex1}
		We consider the dynamics of a parallel RLC circuit with ZIP load (i.e., constant impedance, current and power load)~\citep{kundur1994power}. Its state-space equations are given by
		\begin{align}\label{eq:RLCdyn_ex1}
		-L\dot{I}&=RI+V-u\\
		C\dot{V}&=I-GV-\frac{\bar P}{V}-I_s
		\end{align}
		where $L,C,R,G,\bar P,I_s$ are positive constants, $I(t),V(t)\in \mathbb{R}$ are state-variables, and $u(t)\in \bR$ is the control input. Its total energy is
		\begin{align}
		S(I,V)&=\dfrac{1}{2}LI^2+\dfrac{1}{2}CV^2.
		\end{align}
		The time derivative of the total energy along the trajectories of the system is computed as
		\begin{align*}
		\dot S&= -RI^2-GV^2-\bar P+uI-VI_s
		\leq uI-VI_s.
		\end{align*}
		This implies the system is passive with input $[I,-V]^\top$, and output $[u,I_s]^\top$. However, $I_s$ is constant and cannot be controlled. Therefore, as in Example~1, the standard passivity with total energy $S(I,V)$ is not helpful for designing input $u$. However, it is possible to show that this system is Krasovskii's  passive based on Proposition~\ref{prop:Krasovskii_passivity_BM}. The system can be represented as a gradient system \eqref{BM_gradient_form} as follows
		\begin{align}
		\begin{bmatrix}
		-L & 0\\0&C
		\end{bmatrix}\begin{bmatrix}
		\dot{I}\\ \dot{V}
		\end{bmatrix}=\begin{bmatrix}
		\frac{\partial P}{\partial I}\\\frac{\partial P}{\partial V}
		\end{bmatrix}+\begin{bmatrix}
		1 & 0\\0& -1
		\end{bmatrix}\begin{bmatrix}
		u\\ I_s
		\end{bmatrix}\\
		P=\dfrac{1}{2}RI^2+IV-\dfrac{1}{2}GV^2-\bar P\ln{V}-I_sV,
		\end{align}
		where
		\begin{align*}
		D M  \frac{\partial^2 P }{\partial x^2}+\frac{\partial^2 P }{\partial x^2}MD = \diag \left\{-R, -\left(G-\bar P/V^2\right)\right\}
		\end{align*}
		is negative semidefinite in the set $\mathcal{B}=\{(I,V)\in \mathbb{R}^2|GV^2\geq \bar P\}$ with $M=\diag\{L,C\}$. This implies that the system is Krasovskii's passive for all the trajectories in $\mathcal{B}$ with port-variables $u_d=\dot{u}$ and $\dot{I}$.
	\end{example}
\vspace{-1mm}
\section{Applications of Krasovskii' passivity}
\vspace{-1mm}
\subsection{Primal-dual dynamics}
\vspace{-1mm}
Recently, the primal-dual dynamics corresponding to the convex optimization problem has been studied from the passivity perspective; see, e.g.~\cite{stegink2017unifying}. In this subsection, we reconsider the convex optimization problem from the viewpoint from Krasovskii's passivity.
		
Consider the following equality constrained convex optimization problem
\begin{equation}\label{standard_SOP}
	\begin{aligned}
	&\min_{x \in \mathbb{R}^{n}} F(x)\\
	& \text{subject to} \ h_{i}(x)=0, \ i = 1,\hdots , m,
	\end{aligned}
	\end{equation}
	where $F:\bR^n\rightarrow \bR$ is a class $C^2$ strictly convex function, and $h_i:\bR^n \rightarrow \bR, ~ \forall i\in \{1,\cdots,m\}$ are linear-affine equality constraints. For the sake of simplicity of the notation define $h(x)=[\begin{array}{ccc}h_1 & \cdots & h_m\end{array}]^\top$. We assume that  Slaters condition hold, i.e., there exist an $x$ such that $h(x)=0$. Under this assumption there exist an unique optimum $x^\ast$ to \eqref{standard_SOP}.
	
In the constrained optimization, the Karush-Kuhn-Tucker (KKT) condition gives a necessary condition which also become sufficient under Slaters condition. Define the corresponding Lagrangian function:
\begin{eqnarray}\label{convex_Lagrangian}
	\mathcal{L}(x,\lambda)=F(x)+\sum_{i=1}^{m}\lambda_{i} h_{i}(x).
\end{eqnarray}
Under Slaters condition, $x^\ast$ is an optimal solution to the convex optimization problem if and only if there exist $\lambda_i^\ast\in \mathbb{R}$, $i = 1,\hdots , m$ satisfying the following KKT conditions
	\begin{eqnarray}\label{standard_KKT}
	\frac{\partial \mathcal{L}(x^*,\lambda^*)}{\partial x}=0, \ \frac{\partial \mathcal{L}(x^*,\lambda^*)}{\partial \lambda}=0.
	\end{eqnarray} 
	Since strong duality \citep{boyd2004convex} holds for \eqref{standard_SOP}, $(x^*,\lambda^{*})$ satisfying the KKT conditions \eqref{standard_KKT} is a saddle point of the Lagrangian $\mathcal{L}$. That is, the following holds.
	\begin{equation}
	(x^\ast,\lambda^\ast)=\arg\max_{\lambda}\left(\arg\min_{x}\mathcal{L}(x,\lambda)\right).
	\end{equation}
To seek a saddle point, consider the following so called primal-dual dynamics,
	\begin{equation}\label{maindyn}
	\begin{split}
	-\tau_{x}\dot{x}&=\frac{\partial \mathcal{L}(x,\lambda)}{\partial x} + u,\\
	\tau_{\lambda_{i}}\dot{\lambda}_{i}&=  \frac{\partial \mathcal{L}(x,\lambda)}{\partial \lambda},\hspace{1cm}
	y=x.
	\end{split}
	\end{equation}
where $u, y: \bR \to \mathbb{R}^n$, and $\tau_x\in \bR^{n \times n},~\tau_{\lambda}\in \bR^{m\times m}$ are positive definite constant matrices. If for some constant input $u=u^*$, this system converges to some equilibrium point $(x^*,\lambda^*)$, then this is nothing but a solution to the convex optimization. Indeed, it is possible to show its convergence based on Krasovskii's passivity. 
\begin{proposition}\label{prop:Krasovskii_passivity_PD}
Consider system \eqref{maindyn}. Assume that Slaters conditions hold. Then the primal dual dynamics~\eqref{maindyn} is Krasovskii's passive with respect to a supply rate $h_K:=-\tau_x^{-1}(\partial \mathcal{L}(x,\lambda)/{\partial x}\allowbreak + u)$ for $Q:=\diag\{\tau_x,\tau_{\lambda}\}$. 
\end{proposition}
\begin{pf}
Define
\begin{eqnarray}
	 	g_0:=\begin{bmatrix}
	 	\nabla_{x} F(x)+\sum_{i=1}^{m}\lambda_{i}\nabla_{x} h_{i}(x)\\
	 	h(x).
	 	\end{bmatrix}.
	 \end{eqnarray}
Then $Q_{g_0}$ is computed as %$\diag\{-2\nabla_x^2F(x),0\}$,
	 \begin{align*}
	 			Q_{g_0}(x)
	 			%&=M\dfrac{\partial g_0(x)}{\partial x}+\dfrac{\partial^\top g_0(x)}{\partial x} M\\
	 			&= \begin{bmatrix}
	 			\tau_x&0\\0& \tau_{\lambda}
	 			\end{bmatrix}\begin{bmatrix}
	 			-\tau_x^{-1}\nabla_x^2F(x) & -\tau_x^{-1}\nabla_x^{\top}h(x)\\\tau_{\lambda}^{-1}\nabla_xh(x) &0
	 			\end{bmatrix}\\
	 			&\hspace{5mm}+\begin{bmatrix}
	 			-\tau_x^{-1}\nabla_x^2F(x) & \tau_x^{-1}\nabla_x^{\top}h(x)\\-\tau_{\lambda}^{-1}\nabla_x^{\top}h(x) &0
	 			\end{bmatrix}\begin{bmatrix}
	 			\tau_x&0\\0& \tau_{\lambda}
	 			\end{bmatrix}\\ 
	 			&= 2\begin{bmatrix}
	 			-\nabla_x^2F(x) & 0\\0 &0
	 			\end{bmatrix} \leq 0,
	 \end{align*}
	 i.e., \eqref{eq2:differential_passivity} holds.
	 Moreover, equation \eqref{eq2:differential_passivity} automatically holds for the constant input vector field.\qed
\end{pf}
{Here, we only show Krasovskii's passivity of the primal dual dynamics due the limitation of the space, but one can show the convergence as well by using the storage function as a Lyapunov candidate (see \cite{feijer2010stability}).}
%\red{In \cite{stegink2017unifying}, the authors prove shifted passivity for  \eqref{maindyn}, using which they propose to interconnect them to  physical system as a controller. An implication of Proposition \ref{prop:Krasovskii_passivity_PD} is that, we can also propose Krasovskii's passivity based controller instead of shifted passivity.}
\subsection{Krasovskii's Passivity based Control}
In the previous section, we provide the concept of Krasovskii's passivity and investigate its properties. In this subsection, we provide a control technique based on Krasovskii's passivity, i.e., we generalize a control method shown in Example~\ref{eg:boost}. As demonstrated by the example, our method works for a class of systems for which the standard passivity based control technique may not work.

The fundamental idea in passivity based control (PBC) is achieving passivity of the closed-loop system at the desired operating point. For standard passivity, an idea is to design an appropriate feedback controller which is passive. Since as mentioned, the feedback interconnection of two standard passive systems is again standard passive, and thus the control objective is achieved. In the previous subsection, we have clarified that Krasovskii's passivity is also preserved under the feedback interconnection. Motivated by these results, we consider to design a controller which is passive.

Consider the controller of the form:
\begin{align}\label{eq:cont_dyn}
\begin{split}
	-K_1\dot{\eta} (t)&=K_2\eta(t) -u_c(t)\\
	y_c (t)&=\dot \eta (t) = - K_1^{-1}(K_2\eta(t) -u_c(t))
\end{split}
\end{align}
 where $\eta:\bR \to \bR^{p}$, and~$u_c,~y_c:\bR \to \bR^{m}$ are respectively the state, input and output of the controller. The matrices $K_1,~K_2 \in \bR^{p \times p}$ are symmetric and positive definite.
 
 One can show that the controller \eqref{eq:cont_dyn} is standard passive with respect to the supply-rate $\dot{\eta}^\top u_c$ as follows.
\begin{lemma}\label{lem:cont_dyn_pass}
The controller \eqref{eq:cont_dyn} is passive with respect to the supply-rate $\dot{\eta}^\top u_c$ with the storage function $S_{c}(\eta)=(1/2)\eta^\top K_2\eta$.
\end{lemma}
\begin{pf}
The Lie derivative of $S_{c}$ along the vector fields of \eqref{eq:cont_dyn}, denoted by $\dot S_c$ is
	\begin{align*}
		\dot{S}_c&=\dot{\eta}^\top K_2\eta
		=\dot{\eta}^\top \left(-K_1\dot{\eta}+u_c\right)
		\leq  \dot{\eta}^\top u_c.\qed
	\end{align*}
%That completes the proof. \hfill\qed
\end{pf}

Based on the above lemma, we now propose the following interconnection between the plant and the controller
\begin{align}\label{eq:inter_con_const}
	\begin{bmatrix}
	u_d\\u_c
	\end{bmatrix}=	\begin{bmatrix}
	0 & 1\\-1 &0
	\end{bmatrix}	\begin{bmatrix}
	h_{K}\\y_c
	\end{bmatrix}+\begin{bmatrix}
	0 \\ \nu
	\end{bmatrix}
\end{align}
where $\nu:\bR \to \bR^m$. This interconnection structure is different from that considered in Proposition~\ref{interconnection_kras}. However, it is still possible to conclude that the closed-loop system is disspative by using a similar storage function as Proposition~\ref{interconnection_kras} as follows.
\begin{theorem}\label{cor:Krasovskii_controller}
Consider a system~\eqref{eq:sys_ext} satisfying~\eqref{eq1:differential_passivity} and~\eqref{eq2:differential_passivity} for some symmetric and positive definite matrix~$Q \in \bR^{n\times n}$. Suppose that $\cE$ is not empty, and $(x^*,u^*) \in \cE$ is an isolated equilibrium of the extended system~\eqref{eq:sys_ext}. Also, consider the control dynamics~\eqref{eq:cont_dyn} with $\eta(t) = u^* - u(t)$, $u_c(t) = g^\top(x) Q f(x,u) - \nu (t)$, and $y_c(t) = u_d(t)$. Then, the closed-loop system (consisting of the extended system and controller dynamics) is dissipative with respect to a supply rate $u_d^\top \nu$. Moreover, if $\nu =0$, there exists an open subset~$\bar D \subset \bR^n \times \bR^m$ containing $(x^*,u^*)$ in its interior such that any solution to the closed-loop system starting from~$\bar D$ converges to the largest invariant set contained in
		\begin{align}\label{set:Theorem1_2}
		\{ (x,u) \in \bar D :& \|f(x,u)\|_{Q_{g_0}}=0,\nonumber\\ 
		&K_2(u^* - u) - g^\top(x) Q f(x,u)=0\}.
		\end{align}
\end{theorem}
\begin{pf}
%	\blue{The proof follows by analyzing the closed loop storage function $S_K+S_c$.}
Consider the closed-loop storage function,
	\begin{align}
		S_d(x,u,\eta):=\dfrac{1}{2} \|f(x,u)\|_Q^2+(1/2)\eta^\top K_2\eta. \label{eq:dS_d}
	\end{align}
From Proposition~\ref{prop:Krasovskii_passivity} and Lemma~\ref{lem:cont_dyn_pass}, its Lie derivative along the trajectory of the closed-loop system, denoted by $\dot S_d$, satisfies
\begin{align*}
\hspace{-5mm}\dot S_d = &\|f(x,u)\|_{Q_{g_0}}- \|\dot{\eta}\|_{K_1} + u_d^\top g^\top (x)Q f(x,u) + \dot{\eta}^\top u_c\\
= &\|f(x,u)\|_{Q_{g_0}}- \|\dot{\eta}\|_{K_1}
+u_d^\top (u_c + \nu) - u_d^\top u_c
 \le u_d^\top \nu,
\end{align*}
where from~\eqref{eq1:differential_passivity}, $\|f(x,u)\|_{Q_{g_0}}$ is negative semidefinte at $(x^*,u^*)$.
Then, the closed-loop system is dissipative with respect to a supply rate $u_d^\top \nu$. 

Next, let $\nu =0$. Then, choose $S_d$ as a Lyapunov candidate. Since $(x^*,u^*) \in \cE$ is an isolated equilibrium of the extended system~\eqref{eq:sys_ext}, $S_d(x^*,u^*,0)=0$, and there exists an open subset~$\hat D \subset \bR^n \times \bR^m$ containing $(x^*,u^*)$ in its interior such that $S_d(x,u,\eta) >0$ on $\hat D \setminus\{(x^*,u^*,0)\}$. Therefore, from~\eqref{eq:dS_d} and the LaSalle's invariance principle~[\cite{Khalil:96}], any solution to the closed-loop system starting from~$\hat D$ converges to the largest invariant set contained in
\begin{align*}
&\{ (x,u,\eta) \in \hat D : \|f(x,u)\|_{Q_{g_0}}- \|\dot{\eta}\|_{K_1}=0\}\\
&=\{ (x,u,\eta) \in \hat D : \|f(x,u)\|_{Q_{g_0}}=0, \|\dot{\eta}\|_{K_1}=0\}\\
&=\{ (x,u,\eta) \in \hat D : \|f(x,u)\|_{Q_{g_0}}=0, K_2\eta -u_c=0\}\\
&=\{ (x,u,\eta) \in \hat D : \|f(x,u)\|_{Q_{g_0}}=0,\\
&\hspace{5mm}K_2(u^* - u) - g^\top(x) Q f(x,u)\|=0\},
\end{align*}
where in the first inequality, we use the fact that $K_1$ is positive definite. By considering the projection of the above set on $(x,u)$-space, we obtain the second statement.
\qed
\end{pf}
{This shows us that, if passivity properties of the original system are not easy to find, then one can search for passivity properties of the extended system and design a controller for $u_d$. However, this leads to an dynamics controller for the original system. }
\begin{example}[Revisit of Example \ref{eg:boost}]
Consider the boost converter in Example \ref{eg:boost}. Then, the controller in the form~\eqref{eq:cont_dyn} satisfying the conditions in Theorem~\ref{cor:Krasovskii_controller} is
\begin{align}
\begin{split}
	-K_1 u_d (t)&=K_2 (u^*-u(t)) + (\dot{I}V-\dot{V}I) + \nu (t),\\
	y_c (t)&= u_d(t)
\end{split}
\end{align}
where $\dot I$ and $\dot V$ are used for the sake of notational simplicity. By choosing $K_1=1$ and $\nu=0$, we have the controller~\eqref{eq:con_ex}.
\end{example}
%\balance
\vspace{0mm}
\section{CONCLUSIONS}\label{sec:conclusions}
Inspired by Krasovskii's method for the construction of the Lyapunov function, in this paper, we have proposed the concept of Krasovskii's passivity and studied its properties. Especially, we have shown that Krasovskii's passivity is preserved under the feedback interconnection. This property has been used for Krasovskii's passivity based controller design, which can be designed for a class of systems for which the standard passivity control technique may not be useful. Future work includes to establish bridges with relevant passivity properties such as differential, incremental, shifted passivity properties.

\bibliography{reference}

\end{document}